\documentclass[12pt]{article}
\pdfoutput=1

\usepackage{amssymb,graphicx}
\usepackage[colorlinks=True, linkcolor=violet, citecolor=teal, urlcolor=blue]{hyperref}
\usepackage{cite}
\usepackage{xcolor}
\usepackage[intlimits]{amsmath}
\usepackage{bbm}
\usepackage[small]{subfigure}

\usepackage{MnSymbol}

\makeatletter \@addtoreset{equation}{section} \makeatother

\makeatletter
\let\old@startsection=\@startsection
\let\oldl@section=\l@section
\renewcommand{\@startsection}[6]{\old@startsection{#1}{#2}{#3}{#4}{#5}{#6\mathversion{bold}}}
\renewcommand{\l@section}[2]{\oldl@section{\mathversion{bold}#1}{#2}}
\makeatother

\makeatletter
\let\old@makecaption=\@makecaption
\def\@makecaption{\small\old@makecaption}
\makeatother

\newcommand{\K}{\mathbf{\mathbf{K}}}
\newcommand{\beq}{\begin{equation}}
\newcommand{\eeq}{\end{equation}}
\newcommand{\beqn}{\begin{eqnarray}}
\newcommand{\eeqn}{\end{eqnarray}}
\graphicspath{{figures/}}
\begin{document}



\renewcommand{\thefootnote}{\fnsymbol{footnote}}
\setcounter{footnote}{0}

\begin{center}
{\Large\textbf{\mathversion{bold} Kosterlitz-Thouless transition \\ in two-dimensional superfluid}
\par}

\vspace{0.8cm}

\textrm{Aleksandr Begun${}^{1,2}$, Alexander Molochkov${}^{2,4}$, Konstantin Zarembo${}^{1}$, Anastasiia Zorina${}^{2}$} 
\vspace{4mm}

\textit{${}^1$Nordita, KTH Royal Institute of Technology and Stockholm University, Hannes Alfv\'ens v\"ag 12, 106 91 Stockholm, Sweden}\\
\textit{${}^2$Pacific Quantum Center, Far Eastern Federal University, 10 Ajax settlement, Russkiy Island, 690022, Vladivostok, Russia}\\
\textit{${}^4$Beijing Institute of Mathematical Sciences and Applications, Tsinghua University, 101408, Huairou District, Beijing, China}

\vspace{0.2cm}
\texttt{beg.alex93@gmail.com, molochkov.alexander@gmail.com, zarembo@nordita.org, nastyazorina05@gmail.com}

\vspace{3mm}


\par\vspace{1cm}

\textbf{Abstract} \vspace{3mm}

\begin{minipage}{13cm}
The 2d superfluid (complex $\phi ^4$ theory in  two dimensions) undergoes Kosterlitz-Thouless transition driven by phase fluctuations of the superfluid order parameter. We study the transition by Monte Carlo simulations and also develop an analytic approach based on the effective theory for the Goldstone mode.
\end{minipage}

\end{center}

\vspace{0.5cm}


\setcounter{page}{1}
\renewcommand{\thefootnote}{\arabic{footnote}}
\setcounter{footnote}{0}

\section{Introduction}

Fluctuations of a Goldstone mode radically affect long-distance behavior in two dimensions, giving rise to quasi-long-range order in the phase with nominally broken symmetry   \cite{berezinskii1971destruction,Kosterlitz:1973xp}. The soft mode causes a  slow, algebraic decay of the order parameter 
\begin{equation}\label{quasi-long-range}
 \left\langle \phi (x)\phi (0)\right\rangle\simeq \rho ^2
 \left\langle \,{\rm e}\,^{i\theta (x)}\,{\rm e}\,^{-i\theta (0)}\right\rangle
 =\frac{\,{\rm const}\,}{|x|^\frac{1}{2\pi K^2}}\,,
\end{equation}
consistent with the absence of continuous symmetry breaking
 \cite{Mermin:1966fe,Coleman:1973ci}. 
Here $K$ is the phase stiffness defined through the effective action for the soft mode: 
\begin{equation}\label{XY}
 S=\frac{K}{2}\int_{}^{}d^{2}x\,\left(\partial_{\mu } \theta \right)^{2}.
\end{equation}

This mechanism is at work in many two-dimensional models, notably in 2d QCD with massless quarks  \cite{Witten:1978qu} where gauge interaction produce a non-zero chiral condensate, just like in four dimensions, but in 2d chiral symmetry gets restored by phase decoherence. The  phase stiffness in this model is proportional to the number of colors and in the large-$N$ limit chiral symmetry may appear broken since the correlator (\ref{quasi-long-range}) approaches a constant if the limit of $K\rightarrow \infty $ is taken before $|x|\rightarrow \infty $. Higher orders in $1/N$, properly taken into account, restore the symmetry \cite{Witten:1978qu} as expected on general grounds.

This is true for any number of colors and flavors \cite{Affleck:1985wa}, but in other models phase fluctuations may also generate a gap. The simplest case  is the XY model itself which undergoes the Kosterlitz-Thouless (KT) phase transition \cite{Kosterlitz:1973xp} at
\begin{equation}\label{KT_Kc}
 K_{c}=\frac{2}{\pi }\,.
\end{equation}
The transition is caused by vortices \cite{Minnhagen:1987zz} whose interaction becomes irrelevant at the critical point leading to vortex liberation and a finite correlation length due to Debye screening:
\begin{equation}
 \left\langle \phi (x)\phi (0)\right\rangle\simeq \rho ^2
 \,{\rm e}\,^{-\frac{|x|}{\xi }}\qquad  ({\rm for}~K<K_c).
\end{equation}
The KT transition is of an infinite order with the correlation length scaling exponentially at the critical point \cite{Kosterlitz:1974nba}:
\begin{equation}\label{KT-scaling}
 \xi \sim \,{\rm e}\,^{\frac{\,{\rm const}\,}{\sqrt{K_c-K}}}.
\end{equation}
This peculiar scaling law is the hallmark of the KT transition.

Detecting the KT transition in the XY model is a notoriously difficult task, as known from numerous lattice simulations~\cite{Tobochnik:1979zz,weber1988monte,olsson1991helicity,Gupta:1992zz,ota1992microcanonical,Janke:1993va,schultka1994finite,Olsson:1995kr,lidmar1998dynamical,Hasenbusch:2005xm,komura2012large,hsieh2013finite,Vanderstraeten:2019frg,Jha:2020oik,nguyen2021superfluid}. Often-claimed reasons are the short-distance noise and large finite-size corrections~\cite{weber1988monte,olsson1991helicity,Hasenbusch:2005xm}. 
We will study the KT transition  by Monte Carlo simulations of a slightly different model describing two-dimensional superfluid:
\begin{equation}
 S=\int_{}^{}d^{2}x\,\left(
 \partial _{\mu }\phi ^\dagger \partial ^{\mu }\phi -\mu ^{2}|\phi |^{2}
 +\lambda |\phi |^{4}
 \right).
 \label{eq:cont_action}
\end{equation}
Quite remarkably, the transition appears much sharper and the KT scaling (\ref{KT-scaling}), as we shall see, is well reproduced even on relatively small lattices.

The model is characterized by two parameters, $\lambda $ and $\mu ^2$, both of the  dimension mass-squared. In the simple-minded mean-field approximation the superfluid density is fixed at the potential minimum: 
\begin{equation}
 \left\langle |\phi |^{2}\right\rangle\simeq \frac{\mu ^{2}}{2\lambda }\,.
 \label{eq:mean-field}
\end{equation}
The Goldstone mode (the phase of the condensate) is described by the XY model with the effective stiffness
\begin{equation}\label{phase-stiff}
 K_{\rm MF}=\frac{\mu ^{2}}{\lambda }\,.
\end{equation}
In combination with (\ref{KT_Kc}) this predicts a KT transition at 
\begin{equation}\label{KT-2d}
 \lambda^{\rm MF} _c=\frac{\pi \mu ^2}{2}\,.
\end{equation}

The relation between $\lambda _c$ and $\mu ^2$, as we shall see, is indeed approximately linear, but with a coefficient considerably different from the mean-field prediction. The linear dependence is not at all surprising, it follows from dimensional analysis alone. In the subsequent sections we scrutinize possible deviations from the mean field and present the results of the Monte Carlo simulations.

\section{Beyond mean field}
\label{K-ren}

We will consider a lattice version of the model (\ref{eq:cont_action}):
\begin{equation}\label{lat-phi4}
    S=\sum_{<ij>}|\phi_i-\phi_j|^2+ \sum_i\left( -\mu_0^2|\phi_i|^2+\lambda_0|\phi_i|^4\right),
\end{equation}
where $\mu_0=a\mu,\,\lambda_0=a^2\lambda$ are dimensionless couplings, $a$ being the lattice spacing. The indices $i,j$ label the sites of a square lattice with the periodic boundary conditions, and the summation in the first term is over nearest neighbors. In what follows we make no distinction between $\lambda _0$, $\mu _0$ and $\lambda $, $\mu $, in other words the couplings are measured in the lattice units.

When the modulus of $\phi _i$ is frozen at the minimum of the potential, the effective theory for the phase takes the form of the conventional XY model:
\begin{equation}\label{Lattice-XY}
 S=-K\sum_{\left\langle ij\right\rangle}^{}\cos\left(\theta _i-\theta _j\right),
\end{equation}
where the phase stiffness is given by the same equation~(\ref{phase-stiff}) as in the continuum.  We mention in passing that the lattice spacing cancels out in  (\ref{phase-stiff}), because
 the phase stiffness is dimensionless.

These simple considerations predict (\ref{KT-2d}) for the critical line of the KT transition.  The mean-field estimate however misses two important effects both of which  tend to diminish $\lambda _c$.  The most significant effect is the shift of the transition point induced by small-scale fluctuations of the phase. The critical phase stiffness in the square-lattice XY model differs considerably from~(\ref{KT_Kc}). It cannot be calculated analytically (see nevertheless~\cite{mattis1984transfer}) but is accurately known from numerical simulations. In addition, the mean-field approximation receives corrections of its own. Those are numerically less significant but are responsible for deviations from the linearity of the critical line, clearly visible in the  Monte Carlo simulations. We discuss these two effects in turn.

\subsection{Shift of the critical point}
The qualitative picture of the KT transition as driven by a dilute gas of vortices does not fully reflect the microscopic reality at the lattice. A typical field configuration is a complicated maze of plaquette-size vortices separated by a few lattice steps, strongly interacting with non-topological excitations, the spin waves. The small-scale fluctuations do not change the nature of the transition but strongly renormalize the critical coupling. The shift from the Coulomb-gas value~(\ref{KT_Kc}) is non-universal and depends on how the model is discretized.  For the conventional XY model with the cosine action, the one we are interested in, the critical phase stiffness is accurately known from the Monte Carlo simulations\footnote{An analytic estimate  in \cite{mattis1984transfer} is also in a good agreement with numerics.}. 
Different evaluations, including the most recent ones, 
are compiled in \cite{Jha:2020oik},  and we will use the value quoted there:
\begin{equation}\label{Kclatt}
 K_c^{\rm latt}=1.120\ldots .
\end{equation}
This is almost a factor of two larger than the Coulomb-gas estimate $K_c=2/\pi =0.637\ldots $

The spin waves are apparently important in renormalizing the phase stiffness. This can be seen by comparing to the model with the Villain action. The exact Coulomb gas representation can then be rigorously derived by a duality transformation, while spin waves completely decouple. The critical phase stiffness in the Villain model $K_c^{\rm Villain}=0.752\ldots $  \cite{Janke:1993va} is much closer to the continuum Coulomb-gas estimate.

All in all, a better estimate of the critical $\phi^4$ coupling is $\lambda _c=\mu ^2/K_c^{\rm latt}$. As we shall see, this already gives a good approximation for the transition line but cannot account for small deviations from the linear dependence between $\lambda _c$ and $\mu ^2$. 

\subsection{One-loop correction to phase stiffness}

If the XY  model (\ref{Lattice-XY}) is regarded as the low-energy effective theory of the superfluid, the phase stiffness is a Wilson coefficient obtained by integrating out fast degrees of freedom in the path integral. 
The expression (\ref{phase-stiff}) obtained by freezing the field modulus at the minimum of the potential is obviously an approximation that receives quantum corrections. Those can be systematically accounted for in perturbation theory.
The small parameter is the coupling constant or rather the dimensionless ratio $\lambda /\mu ^2$. The coupling itself is not very small at the critical point ($\lambda_c /\mu ^2\simeq 1/K_c^{\rm latt}\approx 0.9$), but the phase volume of loop integration add a factor of $\sim 1/4\pi $ and we expect quantum correction to be within a ten-percent range. To see if these estimates are correct we will compute the first loop correction explicitly.

 To do so we split the field in the modulus and the phase:
\begin{equation}
 \phi =\frac{\eta }{\sqrt{2}}\,\,{\rm e}\,^{i\theta },
\end{equation}
 and treat the phase as a slow variable, thus doing the path integral in two steps. The phase is frozen in the first step while the modulus is integrated out, with $\theta $ regarded as an external field (we use the continuum notations just for simplicity, loop integrals will be done with lattice propagators):
\begin{equation}
 S=\frac{1}{2}\int_{}^{}d^2x\left[
 \left(\partial \eta \right)^2+\eta ^2\left(\partial \theta \right)^2
 -\mu ^2\eta ^2+\frac{1}{2}\,\lambda \eta ^4
 \right].
\end{equation}
Assuming the phase is slowly varying we treat
\begin{equation}
 v_\mu =\partial_\mu \theta  
\end{equation}
effectively as a constant, to the leading order in the derivative expansion. 

Expanding near the minimum of the potential:
\begin{equation}
 \eta =\sqrt{\frac{\mu ^2-v^2}{\lambda }}+\xi ,
\end{equation}
and integrating out $\xi $, we get in the one-loop approximation:
\begin{equation}
 S_{\rm eff}=-\frac{1}{16\lambda }\int_{}^{}d^2x\,m^4
 +\frac{1}{2}\,\mathop{\mathrm{Tr}}\ln\left(-\partial ^2+m^2\right),
\end{equation}
 with
\begin{equation}
 m^2=2(\mu ^2-v^2).
\end{equation}
Expanding further in $v$ we find:
\begin{equation}
 S_{\rm eff}=\frac{K}{2}\int_{}^{}d^2x\,v^2,
\end{equation}
with the effective stiffness 
\begin{equation}
 K=\frac{\mu ^2}{\lambda }-\int_{-\pi }^{\pi }\frac{d^2p}{(2\pi )^2}\,\,
 \frac{1}{2\mu ^2+\sum\limits_{\mu =1}^{2}(2-2\cos p_\mu )}
\end{equation}
where the integrand is the lattice propagator of a scalar with mass $\sqrt{2}\,\mu$.

The first correction to the phase stiffness is thus given by the self-energy bubble diagram.
To compute the integral we use Schwinger's proper-time representation:
\begin{eqnarray}\label{Kren}
  K&=&\frac{\mu ^2}{\lambda }-\frac{1}{2}\int_{0}^{\infty }dt\,\,{\rm e}\,^{-(2+\mu ^2)t}
  \left(\int_{-\pi }^{\pi }\frac{dp}{2\pi }\,\,\,{\rm e}\,^{t\cos p}\right)^2
  =\frac{\mu ^2}{\lambda }-\frac{1}{2}\int_{0}^{\infty }dt\,\,{\rm e}\,^{-(2+\mu ^2)t}I_0(t)^2
\nonumber \\
&=&
\frac{\mu ^2}{\lambda }-\frac{\K}{(2+\mu ^2)\pi }\,,
\end{eqnarray}
where $I_0(x)$ is the modified Bessel function and $\K$ is the complete elliptic integral of the first kind:
$$\K\equiv \K\left(\frac{2}{2+\mu ^2}\right),$$ 
in the notations of \cite{gradshteyn2014table}.

The one-loop correction is indeed numerically small, and in the range  $\mu ^2>1$ never exceeds $10\%$ of $K_c^{\rm latt}$. An IR divergence at $\mu^2\rightarrow 0$ arises because the Higgs mode becomes light, invalidating the very logic of effective field theory.

\subsection{Improved estimate of the critical coupling}

Taking into account renormalization of the phase stiffness and the explicit one-loop correction (\ref{Kren}), we find for the critical coupling:
\begin{equation}
 \lambda _c=\frac{\mu ^2}{K_c^{\rm latt}+\frac{\K}{\pi (2+\mu ^2)}}\,.
 \label{eq:one_loop}
\end{equation}
with $K_c^{\rm latt}$ given by (\ref{Kclatt}). We use this analytic estimate as a reference point for high-precision Monte Carlo simulations described in the next section.

\section{Monte Carlo results}

\begin{figure}[ht]
    \centering    \includegraphics[width=0.4\textwidth]{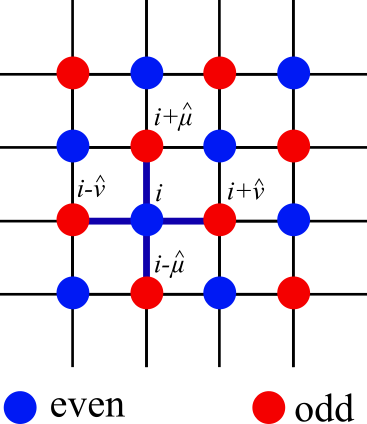}
    \caption{Even-odd labelling of the lattice.}
    \label{fig:evenodd}
\end{figure}

We simulated the model (\ref{lat-phi4}) on the $64^2$ and $128^2$ lattices for a wide range of parameters $\mu^2_0$ and $\lambda_0$ using Markov chain Monte Carlo (MCMC) methods following the standard Metropolis algorithm. To facilitate field sampling we employ an even-odd labelling of the lattice 
(fig. ~\ref{fig:evenodd}). Each site of one sublattice has its nearest neighbours on the other sublattice,  and all sites of each sublattice can be updated simultaneously.  

\begin{figure}[ht]
    \begin{tabular}{c c}
        \includegraphics[width=0.45\linewidth]{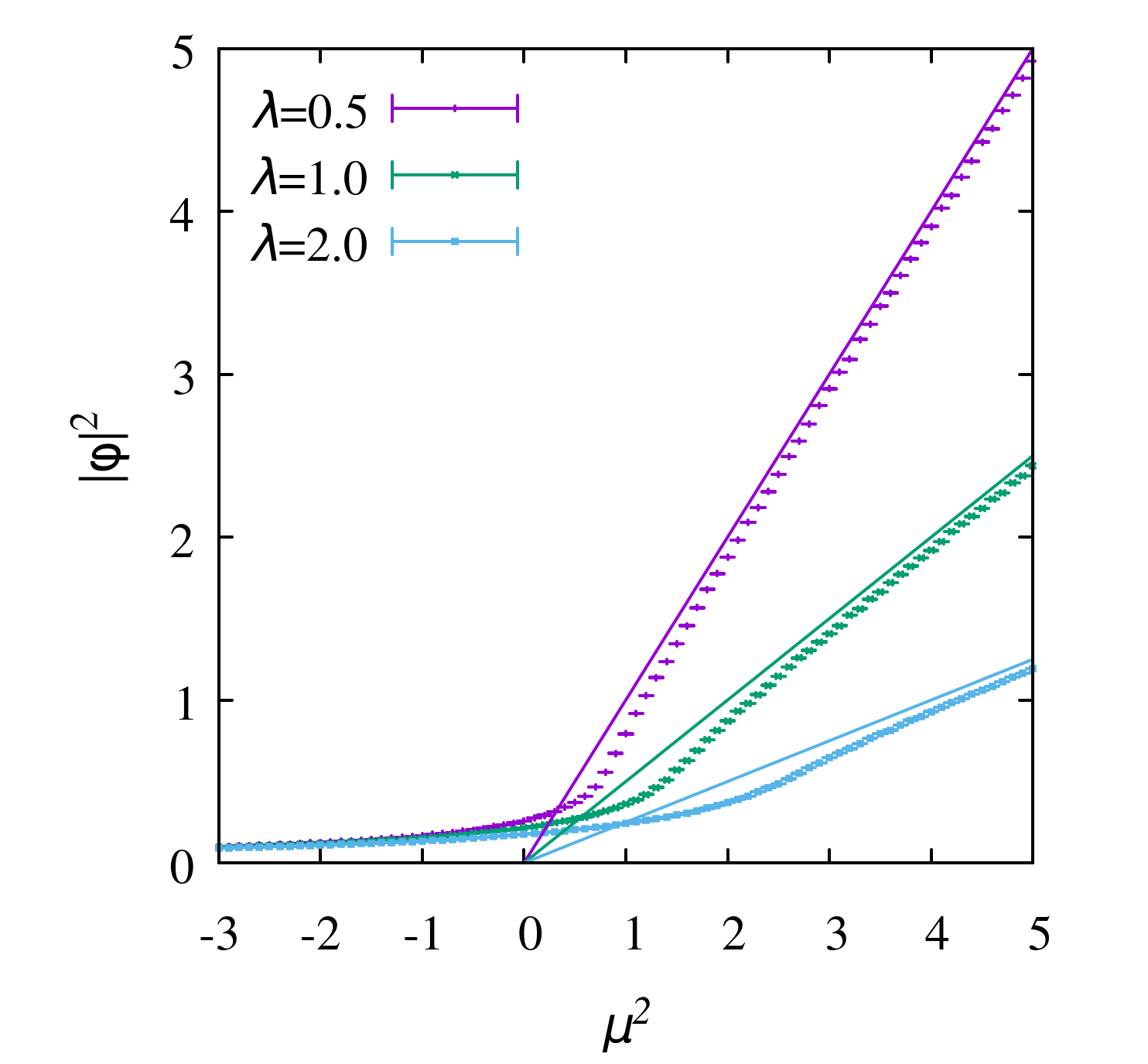} & \includegraphics[width=0.48\linewidth]{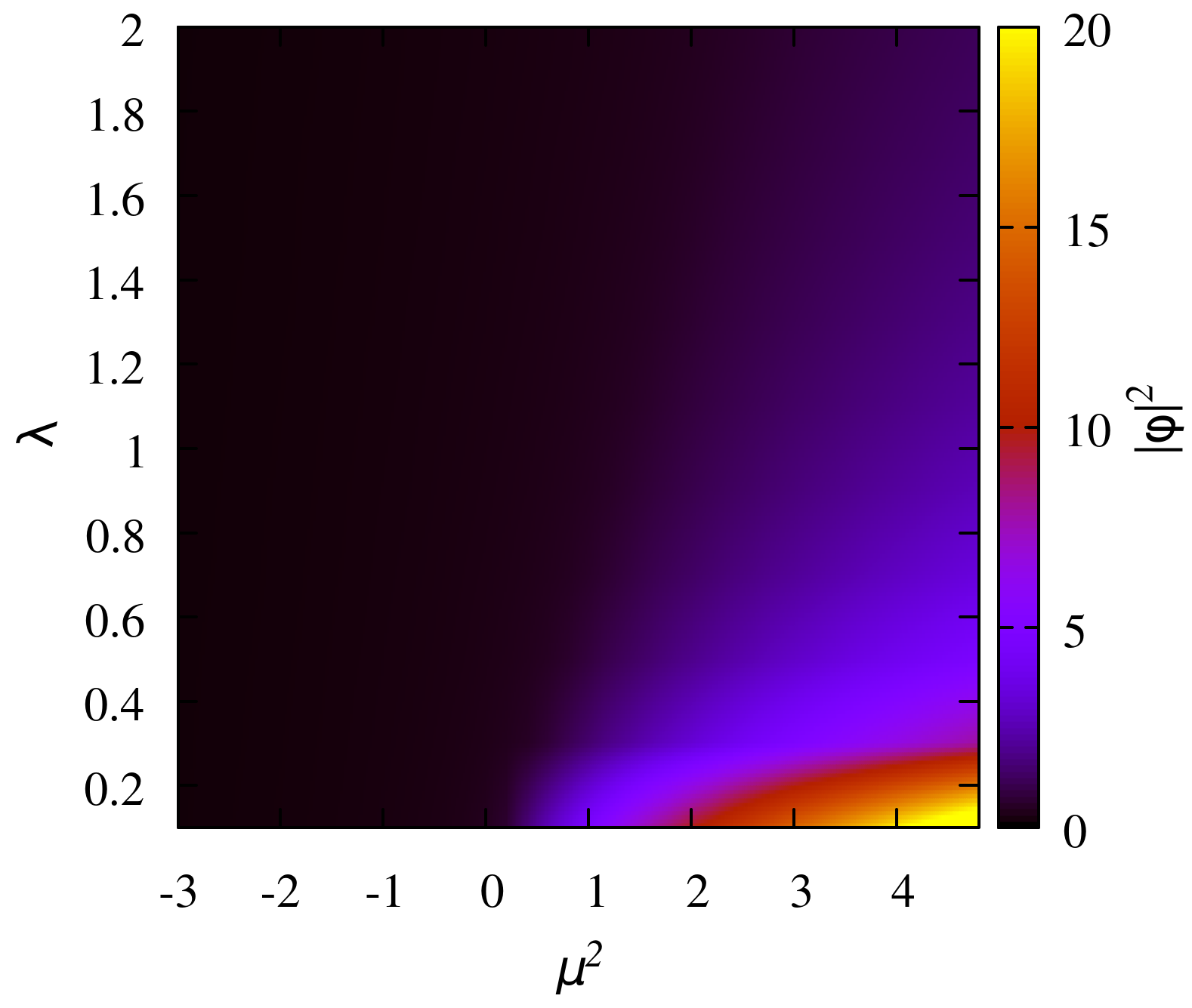} 
    \end{tabular}
    \caption{Expectation value of the square modulus of the field. Straight lines represent the mean-field prediction \eqref{eq:mean-field}.}
    \label{fig:mean-field}
\end{figure}

We measured the superfluid density, the correlation length and the density of vortices for different values of $\mu ^2$ and $\lambda $.
The superfluid density $\langle|\phi|^2\rangle$ is shown in fig.~\ref{fig:mean-field}. From this plot we can get an idea of how accurate the  mean-field approximation is. The simple mean-field estimate  \eqref{eq:mean-field} agrees reasonably well with data in the whole range of parameters and, ss expected, becomes  better with diminishing $\lambda /\mu ^2$.

\subsection{Vortices}

Vortices are topological defects that play a significant role in the behavior of the system, in particular they are responsible for the KT transition. In the context of the $\phi^4$-model, vortices are essentially regions where the phase of the field exhibits a nontrivial circulation pattern.
They are topologically stable and cannot be smoothly deformed into the uniform background without changing the topology. 

\begin{figure}[th]
    \centering    \includegraphics[width=0.8\linewidth]{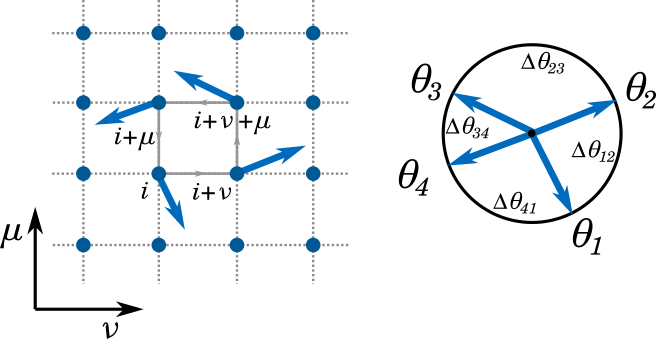}
    \caption{A vortex on the square lattice. The arrows indicate the direction of the field in the complex plane.}
    \label{fig:vortex}
\end{figure}

Elementary vortices can be defined locally, on each plaquette. Consider a plaquette ancored to the $i$-th site and built with links pointing to $i+\hat\nu$ and $i+\hat\mu$. Label the phase differences of the field on the four links of the plaquette by $\Delta \theta _{ab}$. The phase difference is ambiguous $\mathop{\mathrm{mod}} 2\pi $ and we define it by restricting to the fundamental domain $-\pi <\Delta \theta _{ab}<\pi $.
Vorticity is the total phase increment as we go around the plaquette in the counterclockwise direction:
\beq
v=\frac{\sum\limits_i\Delta \theta_{i+1,i}}{2\pi}
\eeq
 Each phase difference is the shortest arch on the unit circle between the two complex values of the field (fig.~\ref{fig:vortex}). Vorticity can clearly take values $-1,0,1$.
Positive $v$ indicates a vortex inside the plaquette, negative $v$ indicates an antivortex.

\begin{figure}[th]
    \centering
    \begin{tabular}{c c}
        \includegraphics[width=0.43\linewidth]{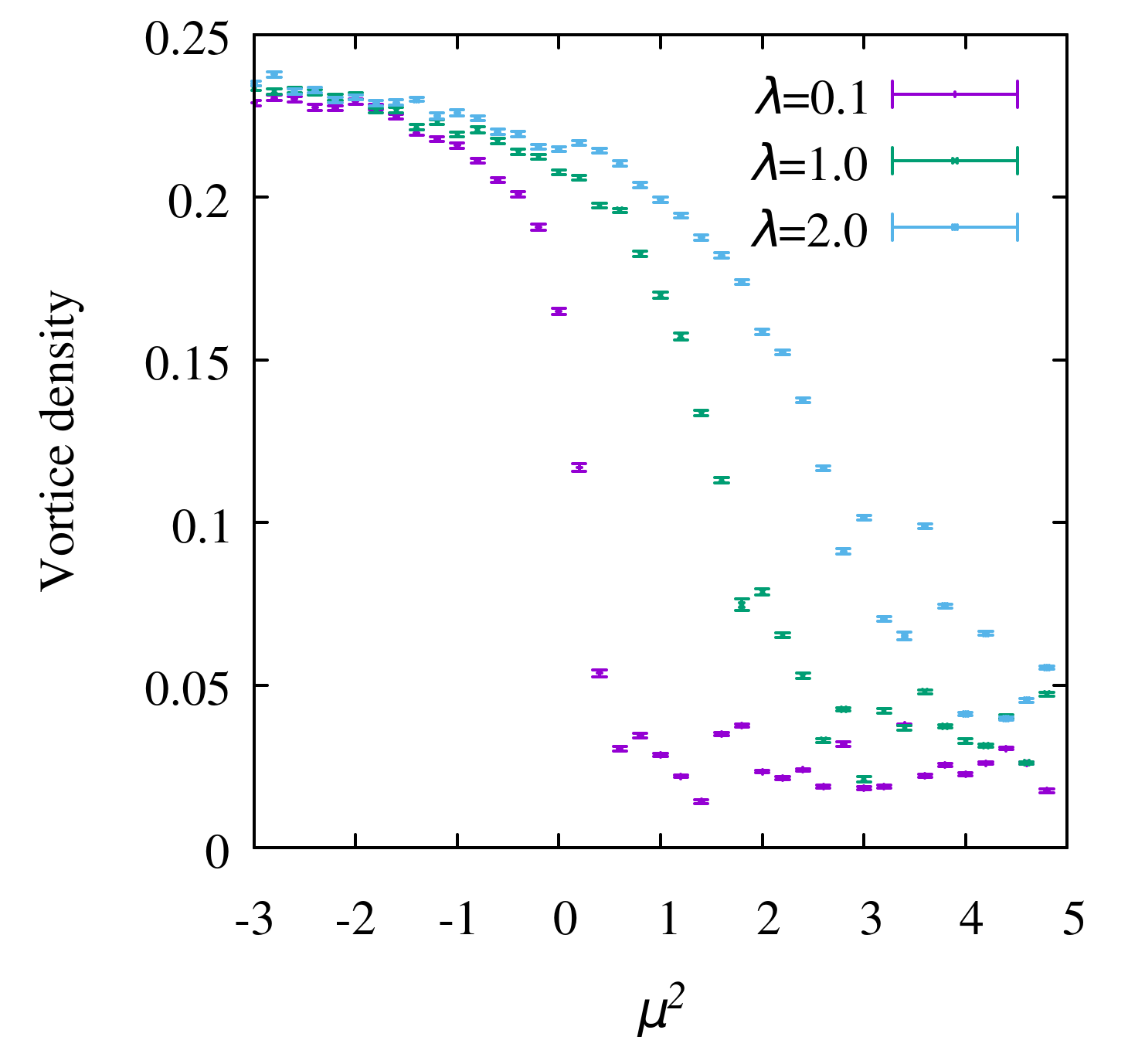} & \includegraphics[width=0.49\linewidth]{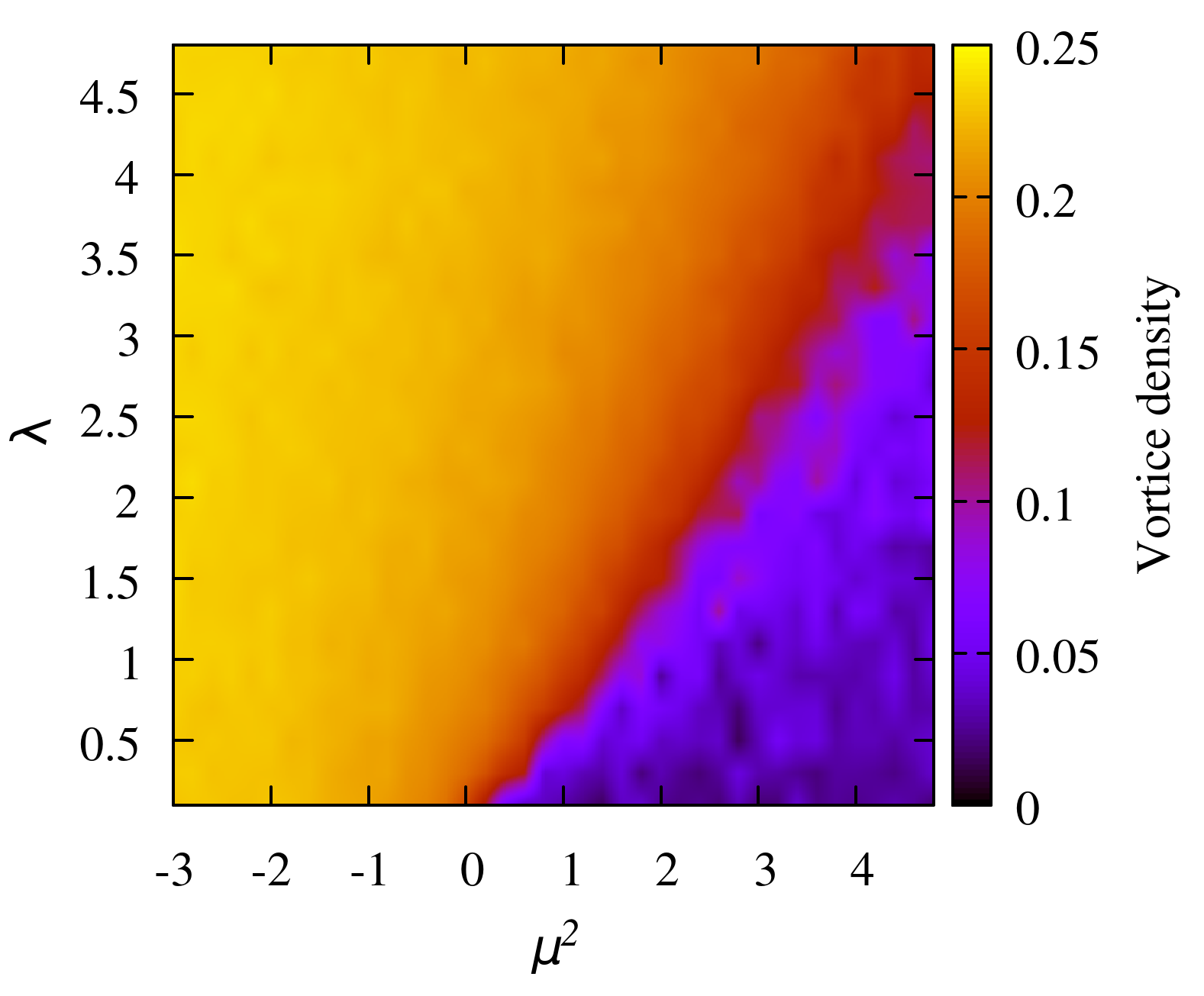} 
    \end{tabular}
    \caption{Vortex density on the $64^2$ lattice.}
    \label{fig:vort-dens}
\end{figure}

\begin{figure}[th]
    \begin{tabular}{c c}
        \includegraphics[width=0.46\textwidth]{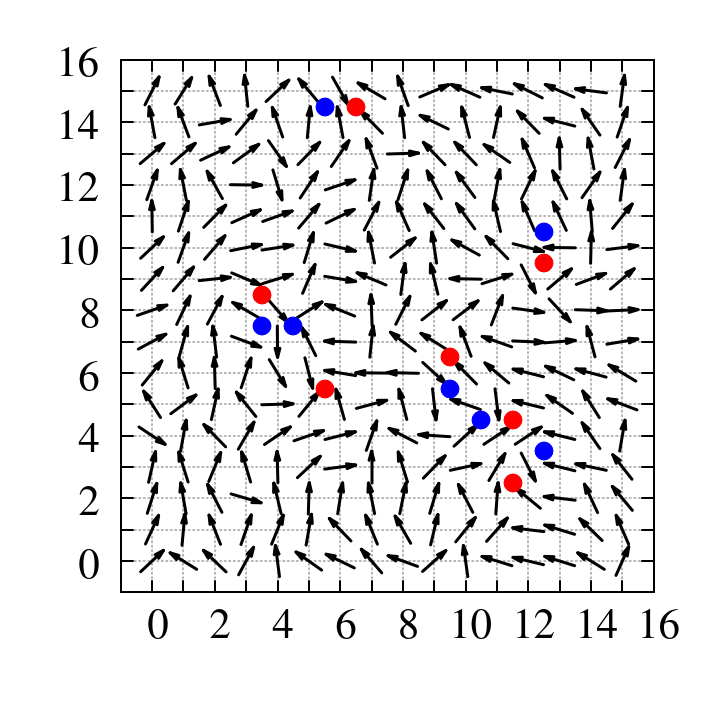}&\includegraphics[width=0.46\textwidth]{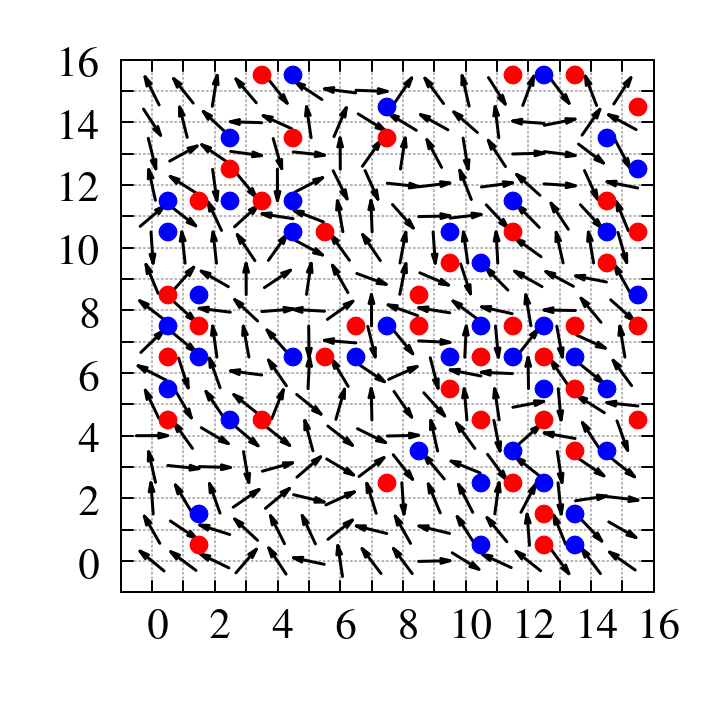} 
    \end{tabular}
    \caption{Vortices in lattice configurations in different regimes ($\mu^2=1.45$, $\lambda=0.80$ and $\mu^2=0.45$, $\lambda=0.80$ left and right panels correspondingly). Arrows represent the complex field phase. Vortices are denoted by red circles, and antivotices are denoted by blue circles. In the dilute regime vortices are bound in pairs  by Coulomb interaction, while in the dense phase they deconfine and percolate the whole lattice.}
    \label{fig:confs}
\end{figure}

The density of vortices dimishes with $\mu ^2$ (fig.~\ref{fig:vort-dens}), however vorticity undergoes only a smooth crossover from the dense to dilute regimes, typical field configurations in each case are illustrated in fig.~\ref{fig:confs}.

\subsection{Correlation length and KT transition}

\begin{figure}[th]
    \centering
   \begin{tabular}{c c}
        \includegraphics[width=0.46\textwidth]{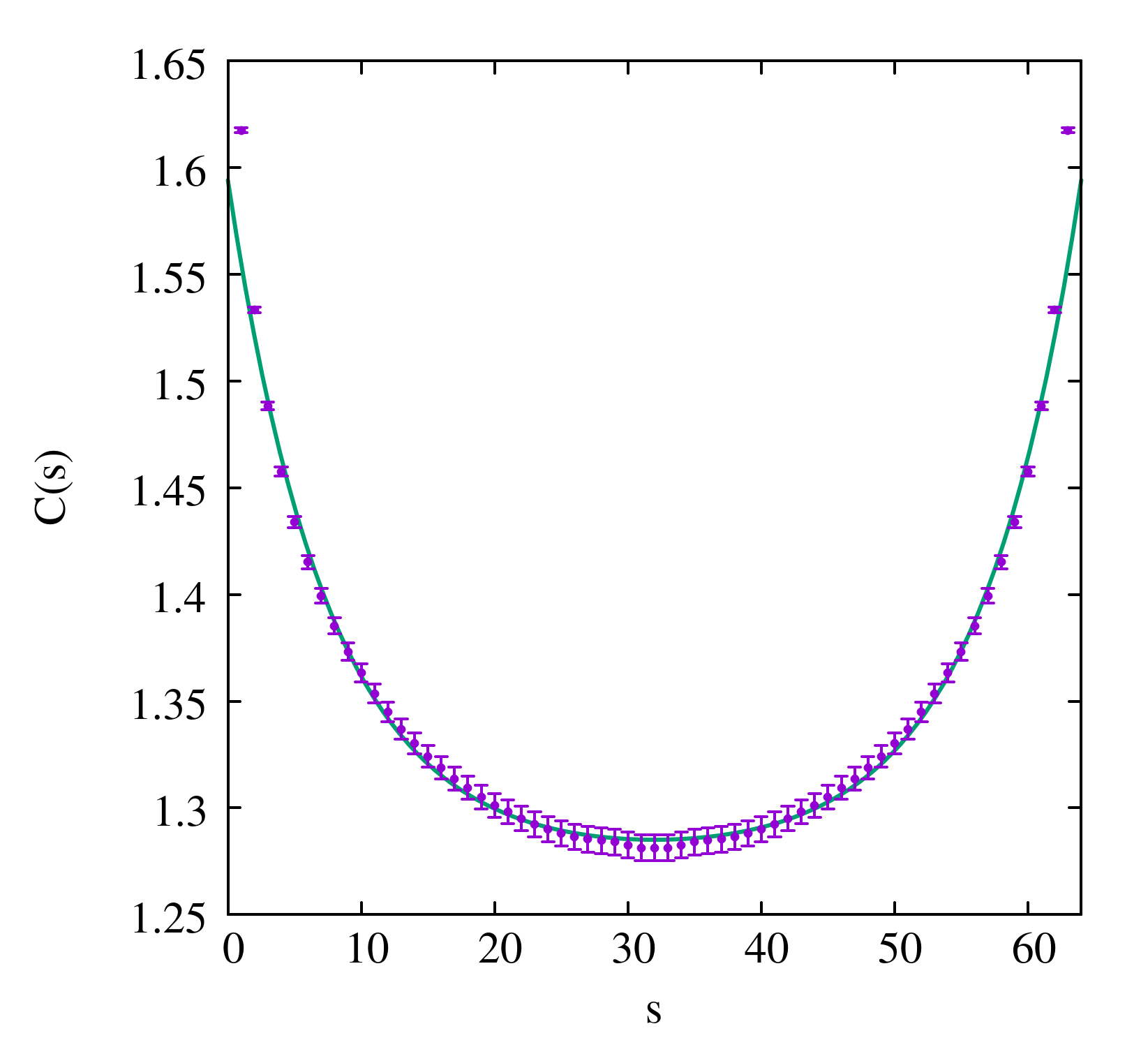}&\includegraphics[width=0.46\textwidth]{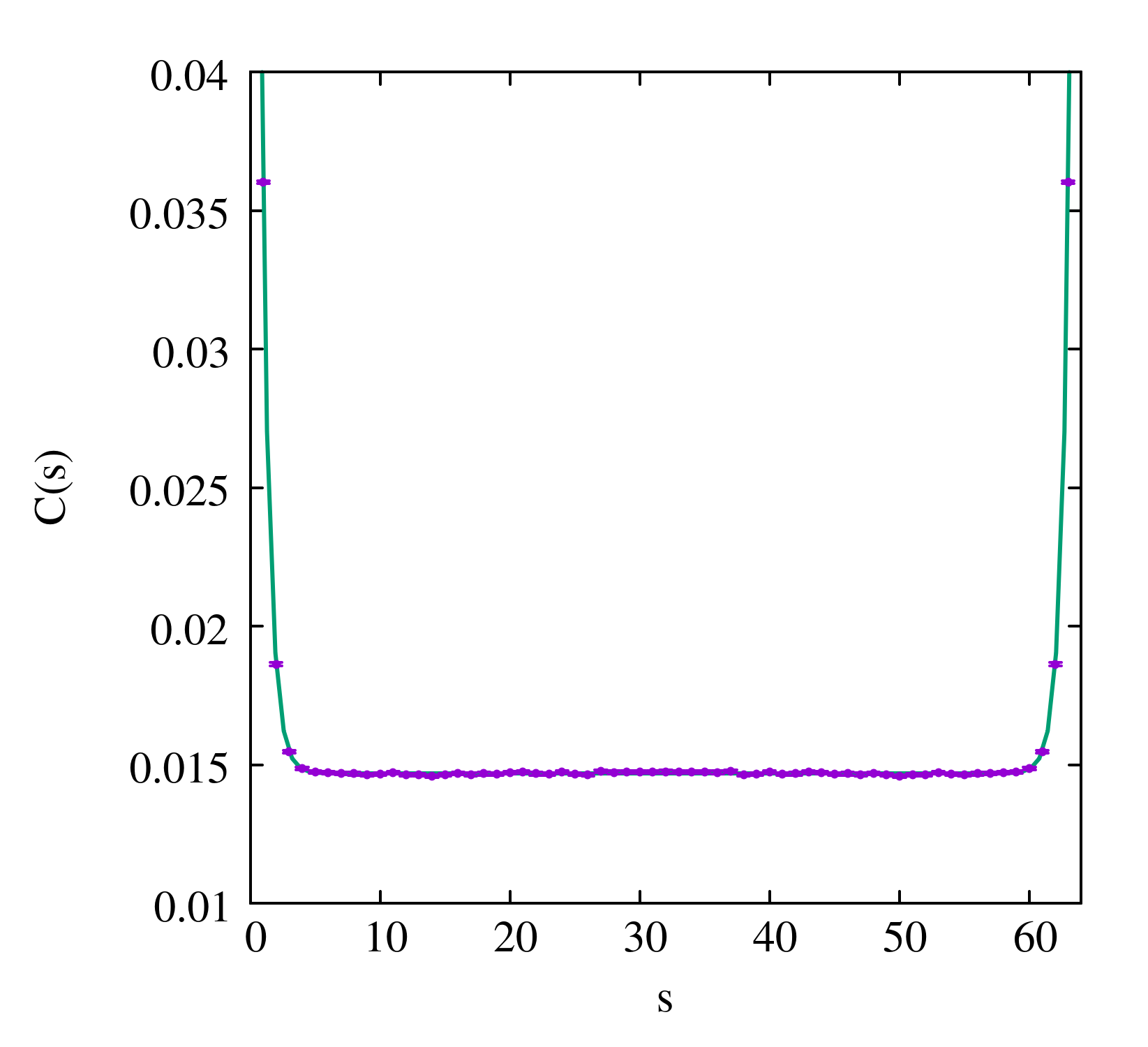} 
    \end{tabular}
    \caption{Correlation function in different KT-phases. The left panel stands for the paired ($\mu^2=1.20$, $\lambda=0.60$) and the right panel for the unpaired phase ($\mu^2=0.45$, $\lambda=0.60$).}
    \label{fig:correlators}
\end{figure}

The density of vortices in itself is not a good indicator of the KT transition. To study the crtitical behavior we consider instead the scalar field correlator:
\beq
C(s)=\sum_{i,\hat\nu}\frac{\eta _i\eta _{i+s\hat\nu}\cos(\theta_i-\theta_{i+s\hat\nu})}{2N_sN_t}.
\eeq
Its typical behavior, shown in fig.~\ref{fig:correlators}, is consistent with the exponential decay at low $\mu ^2$, large $\lambda $,  but not in the large-$\mu ^2$, low $\lambda $ regime where the correlator decays much slower. This is not unexpected since at large-$\mu ^2$ the effective phase stiffness (\ref{phase-stiff}) is big, vortices are bound in pairs and the correlator is expected to decay algebraically, while at low $\mu^2$ vortices are unbound and correlations are screened on the scale of the Debye length.

\begin{figure}[th!]
    \centering
   \begin{tabular}{c c}
        \includegraphics[width=0.46\textwidth]{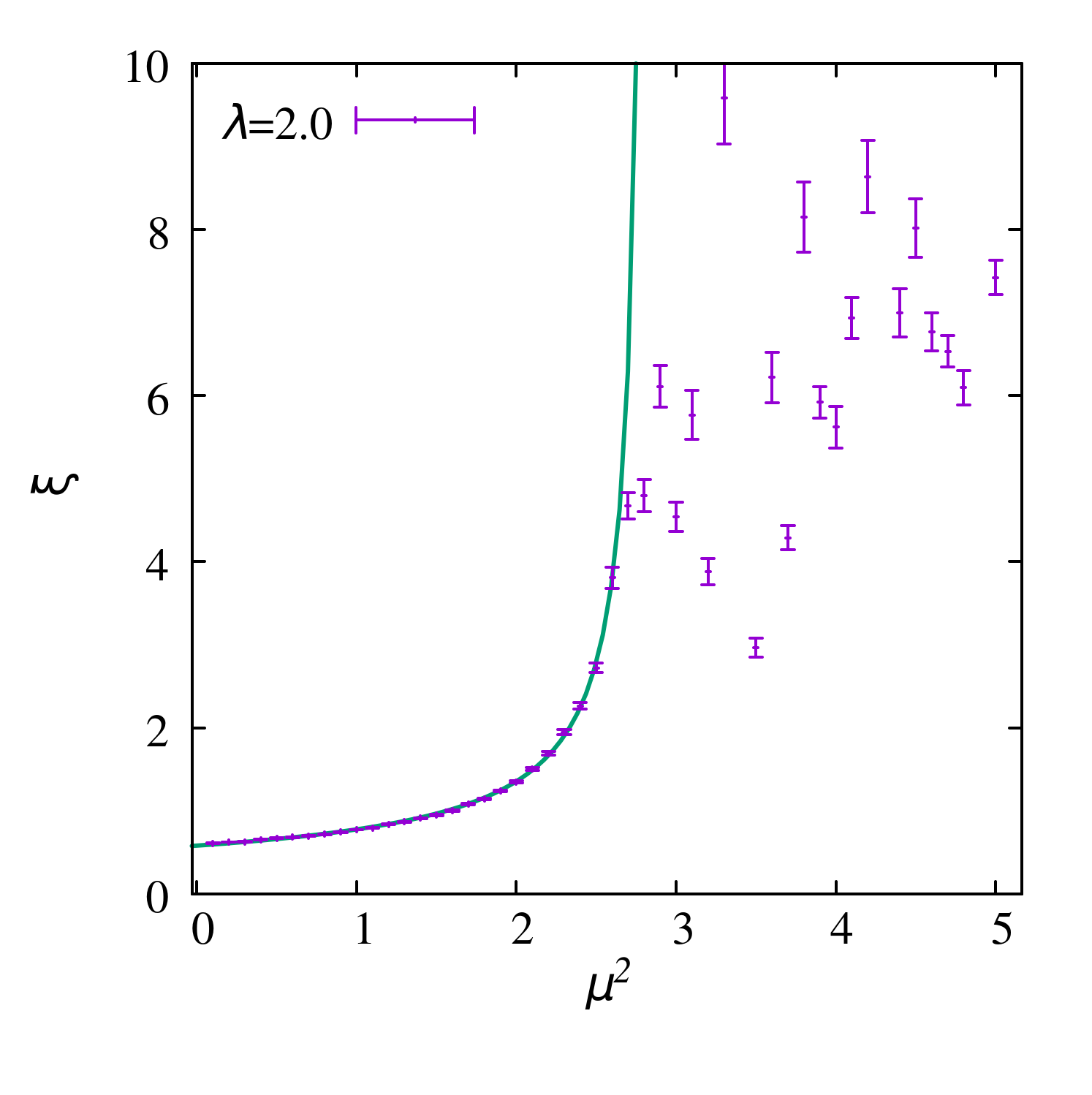}&\includegraphics[width=0.46\textwidth]{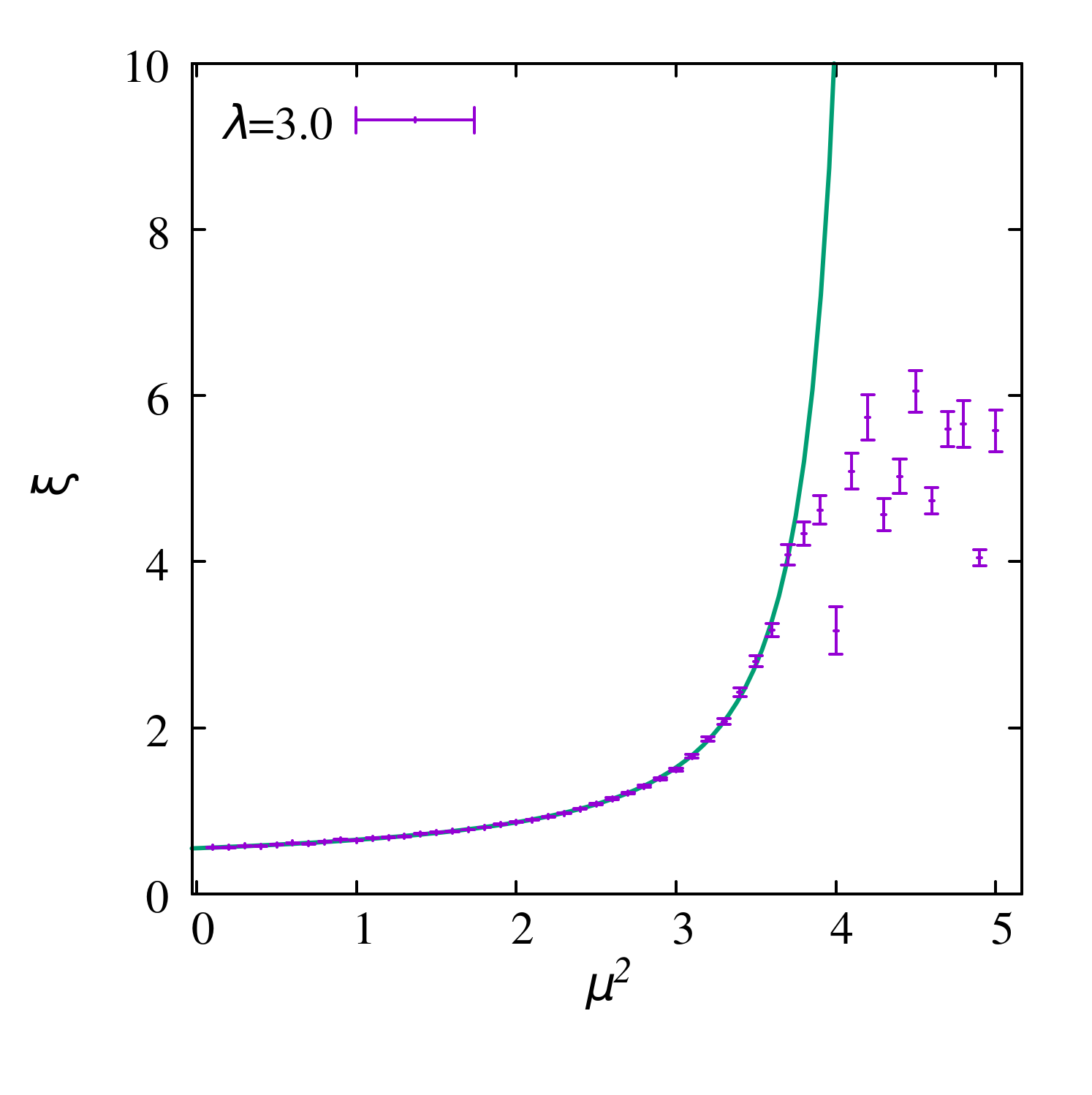} 
    \end{tabular}
    \caption{The correlation length on the lattice of size $64^2$ at different values of $\lambda$: $\lambda=2.0$ in the left panel and $\lambda=3.0$ in the right panel.
    The green lines represent the KT scaling (\ref{eq:crit_exp}).}
    \label{fig:critical_exponent}
\end{figure}

\begin{figure}[th!]
    \centering
   \begin{tabular}{c c}
        \includegraphics[width=0.46\textwidth]{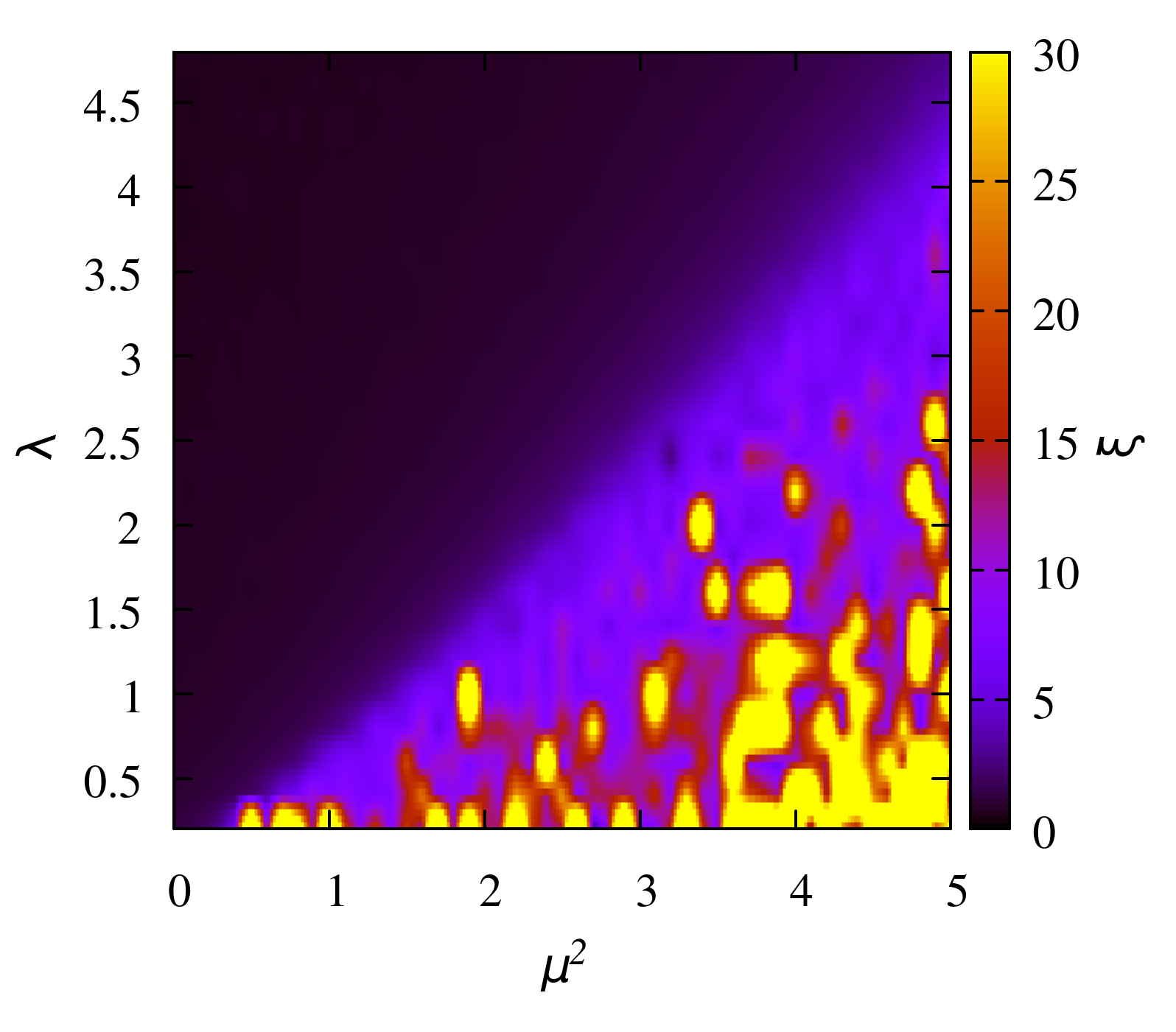}&\includegraphics[width=0.46\textwidth]{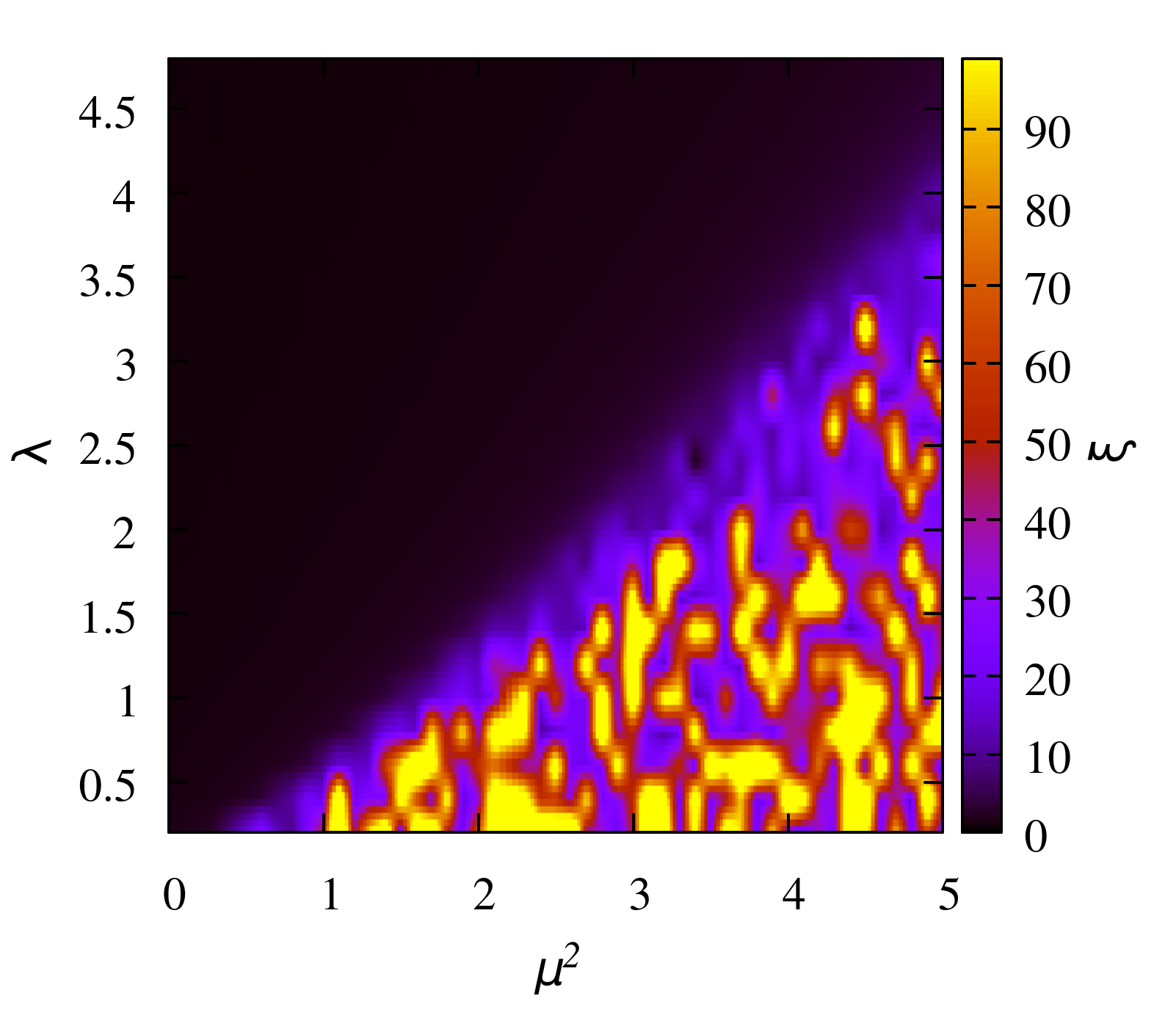} 
    \end{tabular}
    \caption{The correlation length on the lattice of size $64^2$ (left panel) and $128^2$ (right panel).}
        \label{fig:corlength}
\end{figure}

We fitted the correlator with the function
\beq
C(s)=a \left(e^{-\frac{L-s}{\xi}}+e^{-\frac{s}{\xi}}\right)+b.
\label{eq:correl}
\eeq
The correlation length extracted from the fit is shown in fig.~ \ref{fig:critical_exponent}. It grows with $\mu ^2$ as expected, in fact its behavior agrees perfectly well with the KT scaling:
\beqn
\label{eq:crit_exp}
\xi \sim & \,{\rm e}\,^{\frac{\mathrm{const}}{\sqrt{\mu_{crit}^2-\mu^2}}} .
\eeqn
When $\mu ^2$ exceeds $\mu ^2_{crit}$, the quality of fit drops. This simply means that (\ref{eq:correl}) is no longer a good approximation to the correlation function. The correlation length is zero in the low-temperature phase and an exponential fit to an algebraically decaying correlator produces  a "random scatter plot". The phase transition line is clearly visible in fig.~\ref{fig:corlength} where we plot the correlation length as a function of $\mu ^2$ and $\lambda $.

\begin{figure}[th!]
    \centering
   \begin{tabular}{c c}
        \includegraphics[width=0.46\textwidth]{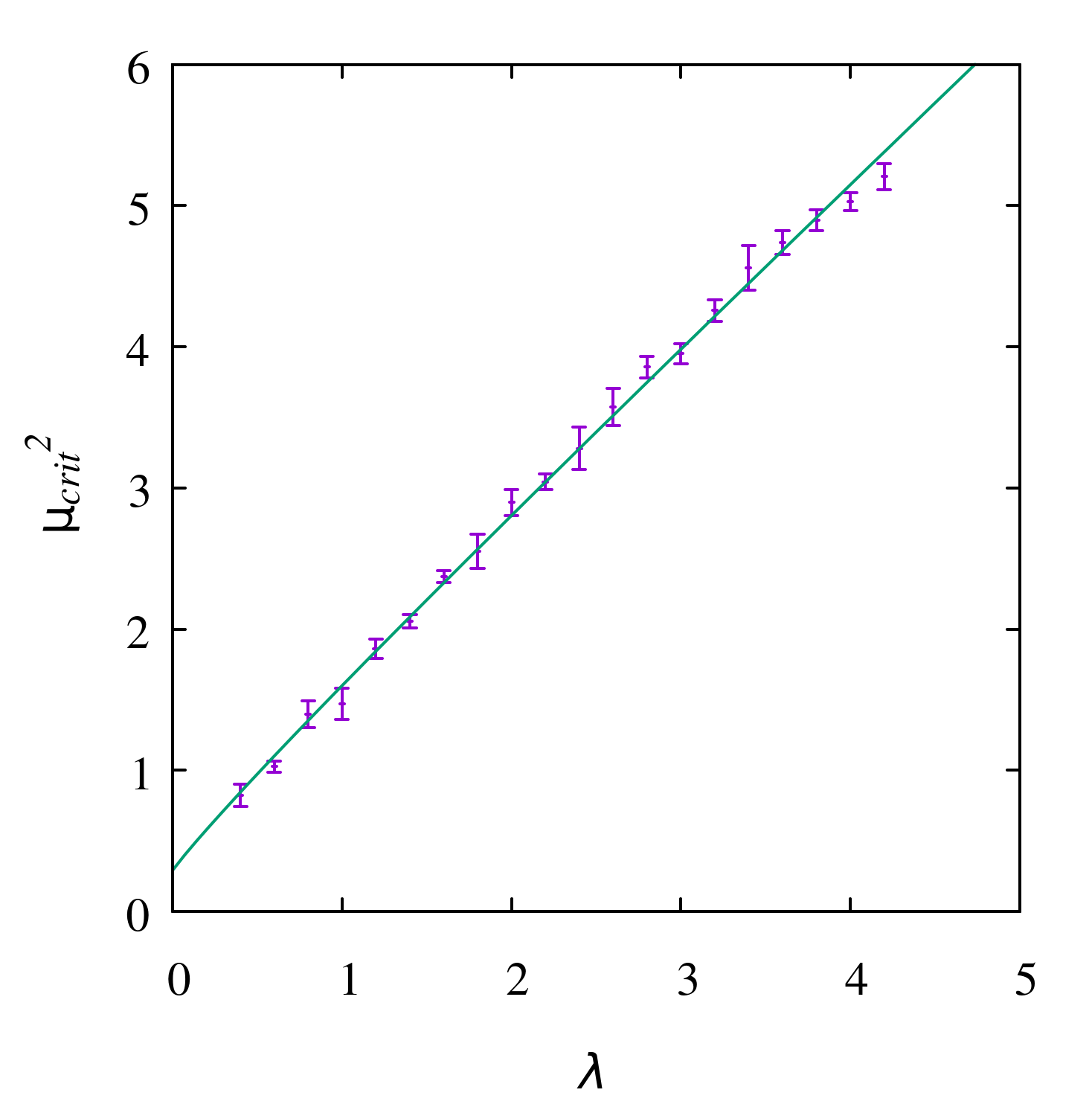}&\includegraphics[width=0.46\textwidth]{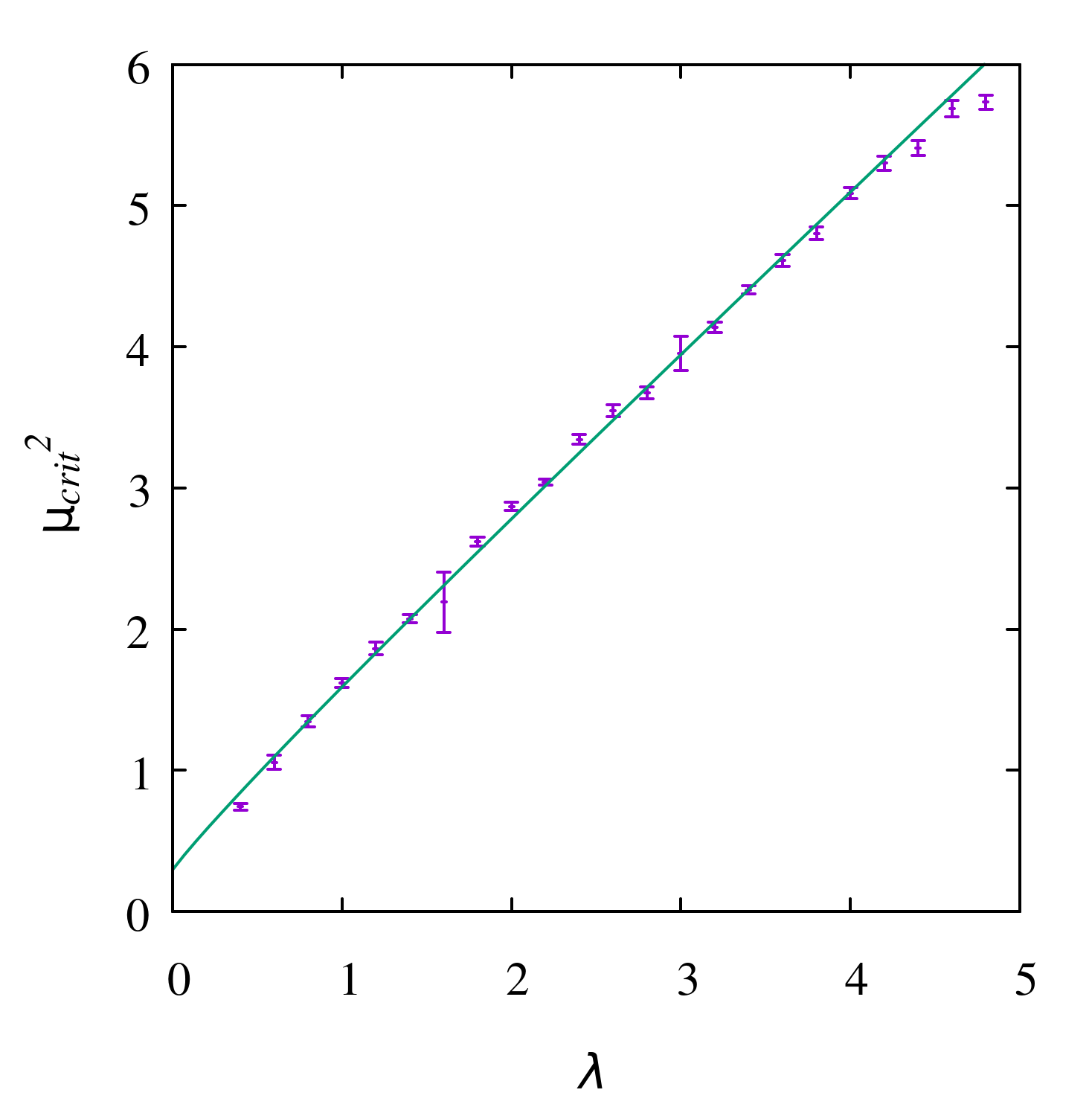} \\
                \includegraphics[width=0.46\textwidth]{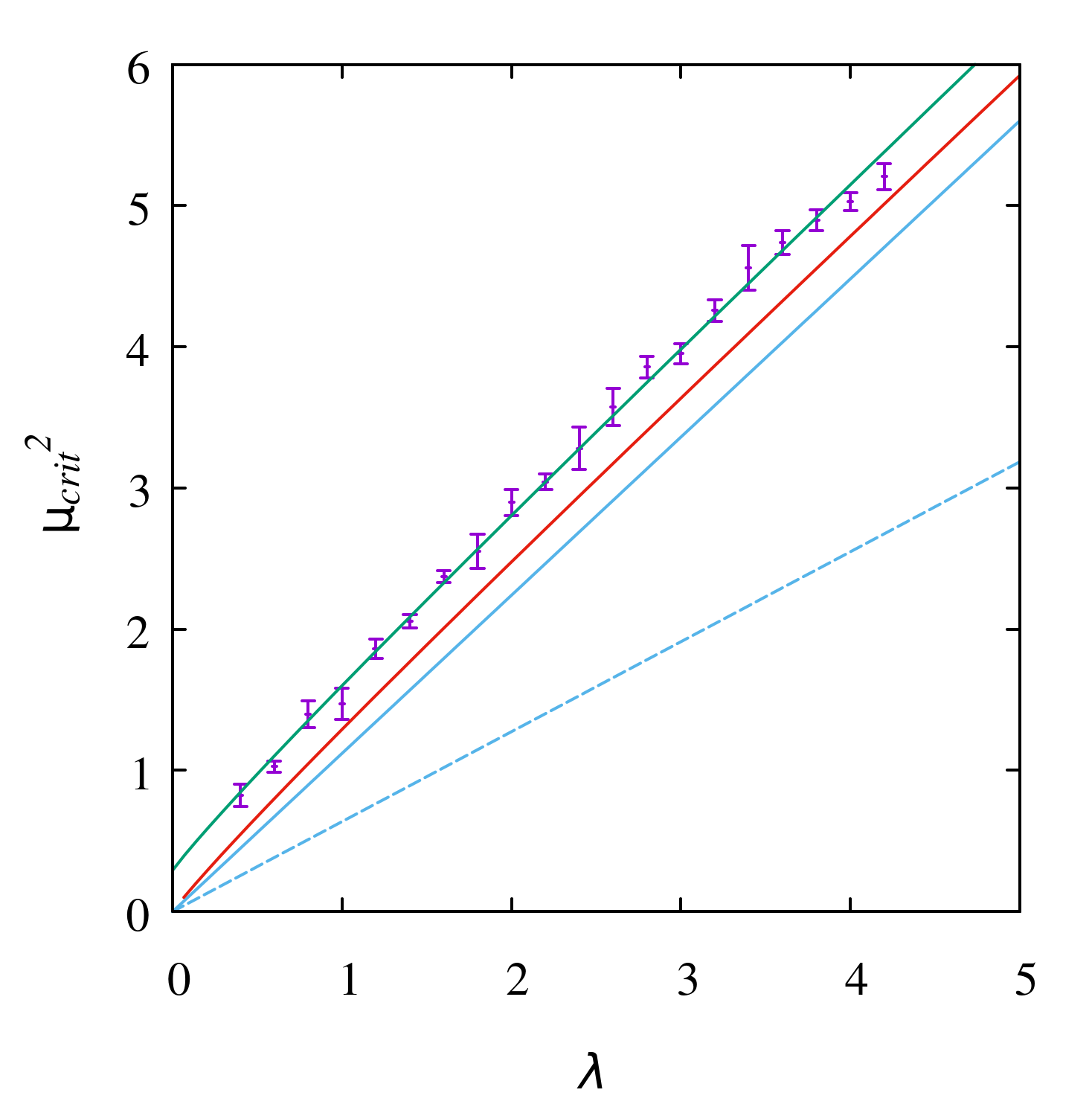}&\includegraphics[width=0.46\textwidth]{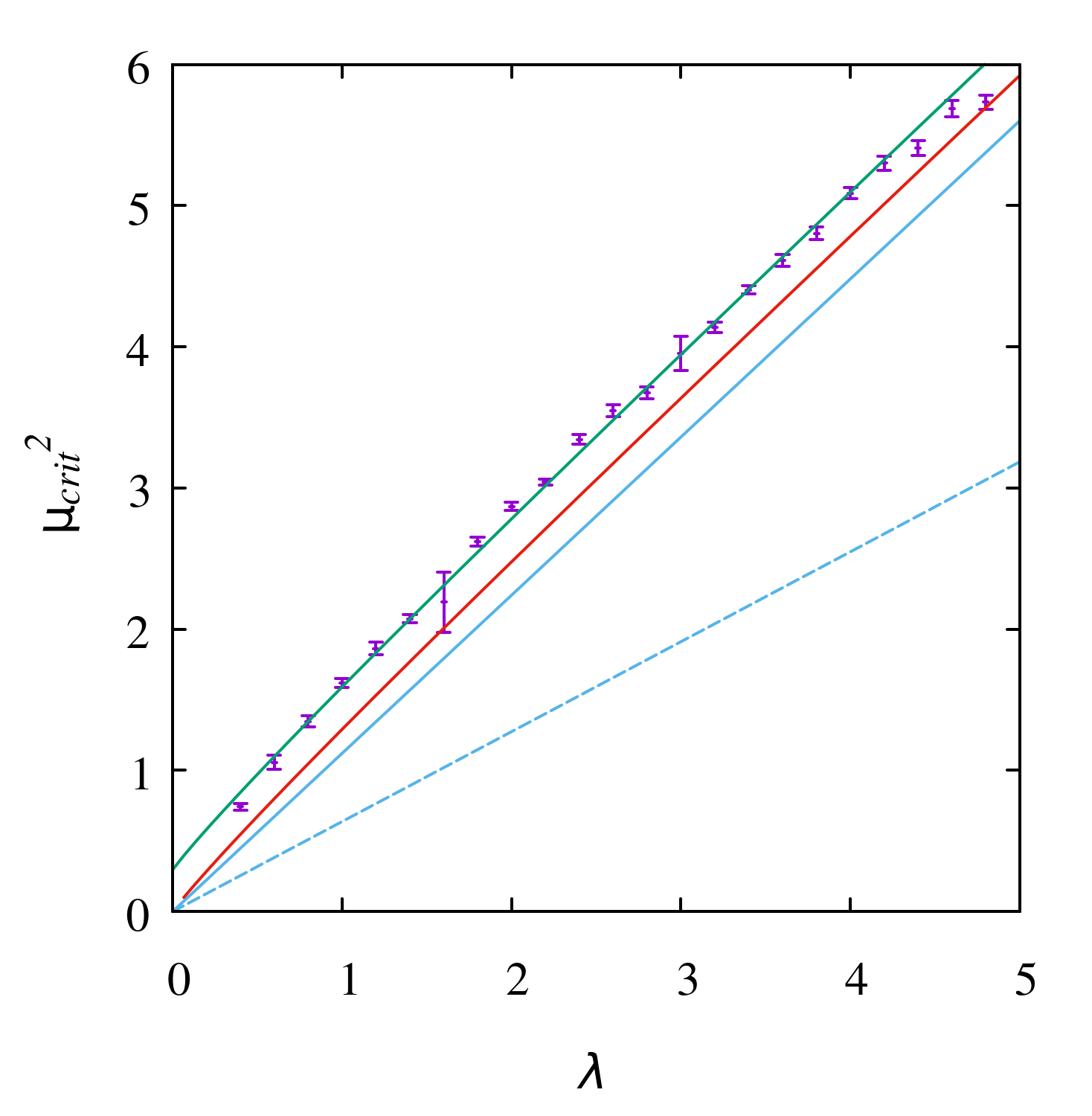} 
    \end{tabular}
    \caption{The phase diagram of the $\phi^4$ model. The left panels are for the $64^2$ lattice, the right ones for $128^2$. The data points are obtained by fitting the correlation length to (\ref{eq:crit_exp}) for various values of $\lambda $.
     The green line is \eqref{eq:fit_f} with the best-fit parameters given in the table. In the lower panel we compare to analytical estimates from sec.~\ref{K-ren}: The dashed blue line is the naive mean field approximation \eqref{KT-2d}, we see that it is not a good approximation to the data. The solid blue line takes into account the shift of the critical point \eqref{Kclatt}, but uses the tree-level result for the effective phase stiffness (\ref{phase-stiff}). It is already much closer to the true data. The red line takes into account the one loop correction \eqref{eq:one_loop}.}
    \label{fig:critical_line}
\end{figure}

The phase diagram of the model is shown in fig.~\ref{fig:critical_line}. Inspired by the improved mean-field approximation~(\ref{Kclatt}), we parameterize the critical line as
\begin{equation}
 \lambda _c=\frac{\mu ^2+d}{K_{crit}+\frac{\K}{\pi (2+\mu ^2+d)}}\,,
 \label{eq:fit_f}
\end{equation}
where the argument of the elliptic $\K$ is $2/(2+\mu ^2+d)$. Here $K_{crit}$ and $d$ are the fit parameters whose best-fit values are given in the table:
\begin{table}[th!]
    \centering
    \begin{tabular}{|c|c|c|}
    \hline
    $L$   & $K_{crit}$ & $d$\\
    \hline
    64  & $1.139\pm 0.016$& $-0.29\pm0.04$\\
    128 & $1.124\pm 0.012$ & $-0.29\pm0.03$\\
    \hline
    \end{tabular}
    \label{tab:my_label}
\end{table}

The parameter $d$ is a loose substitute for unaccounted higher-loop corrections. The critical value of the phase stiffness extracted from our data agrees well with the one in the XY model, see~(\ref{Kclatt}); the XY model can thus be used as an effective theory for the KT transition in the 2d superfluid even quantitatively.

\section{Conclusions}

Our simulations were done on relatively small lattices, and yet the KT transition was clearly visible, with perfect critical scaling of the correlation length. We take this as an indication that finite-size effects in the $\phi^4$ theory are milder than in the pure XY model. It would be interesting to understand why. 

The XY model can be regarded as an effective field theory arising upon integrating out local fluctuations of the superfluid density. We outlined how the effective theory can be constructed within systematic perturbation theory, which led to reasonably accurate analytic predictions for critical behavior. It was, of course, clear from the outset that in the simple model we studied, the phase fluctuations are responsible for the KT transition, making it easy to identify the right low-energy variables. However, we believe that the whole approach can have wider significance and may be applicable to more complicated systems where accurate Monte Carlo simulations are prohibitively difficult, for instance, due to the fermion sign problem.

\subsection*{Acknowledgements}

AB is supported by the Carl Trygger Foundation Grant No. CTS 18:276. 

\appendix

\bibliographystyle{nb-titles}
\bibliography{refs}

\end{document}